\documentclass[aps,preprint,prb,showpacs,amsmath,superscriptaddress,onecolumn,floatfix,nobalancelastpage]{revtex4-1}

\usepackage[nottoc,numbib]{tocbibind}

\usepackage[colorlinks=true, citecolor=black, linkcolor=black, urlcolor=black]{hyperref}

\usepackage[all]{hypcap}
\usepackage{amsmath}
\usepackage{enumerate,bm}
\usepackage{graphicx}
\usepackage{grffile} 
\usepackage{color}
\usepackage{esint}
\usepackage{textpos}
\usepackage[usenames,dvipsnames]{xcolor}
\usepackage{subfigure}
\usepackage{flushend}
\usepackage{tabularx}
\usepackage{amssymb}
\usepackage{float}
\usepackage{subfigure}
\usepackage{ulem} 

\begin{document}


\author{Hannes~H\"ubener}
\affiliation{Nano-Bio Spectroscopy Group and ETSF, Universidad del Pa\'is Vasco, CFM CSIC-UPV/EHU, 20018 San Sebasti\'an, Spain}

\author{Michael~A.~Sentef}
\affiliation{Max Planck Institute for the Structure and Dynamics of Matter and Center for Free Electron Laser Science, 22761 Hamburg, Germany}

\author{Umberto~De~Giovannini}
\affiliation{Nano-Bio Spectroscopy Group and ETSF, Universidad del Pa\'is Vasco, CFM CSIC-UPV/EHU, 20018 San Sebasti\'an, Spain}

\author{Alexander~F.~Kemper}
\affiliation{Department of Physics, North Carolina State University, Raleigh, NC 27695-8202, USA}

\author{Angel~Rubio}
\affiliation{Nano-Bio Spectroscopy Group and ETSF, Universidad del Pa\'is Vasco, CFM CSIC-UPV/EHU, 20018 San Sebasti\'an, Spain}
\affiliation{Max Planck Institute for the Structure and Dynamics of Matter and
Center for Free Electron Laser Science, 22761 Hamburg, Germany}


\title{
Creating stable Floquet-Weyl semimetals by laser-driving of 3D Dirac materials
}
\date{\today}
\begin{abstract}
Tuning and stabilising topological states, such as Weyl semimetals, Dirac semimetals, or topological insulators, is emerging as one of the major topics in materials science. Periodic driving of many-body systems offers a platform to design Floquet states of matter with tunable electronic properties on ultrafast time scales. Here we show by first principles calculations how femtosecond laser pulses with circularly polarised light can be used to switch between Weyl semimetal, Dirac semimetal, and topological insulator states in a prototypical 3D Dirac material, Na$_3$Bi. Our findings are general and apply to any 3D Dirac semimetal. We discuss the concept of time-dependent bands and steering of Floquet-Weyl points (Floquet-WPs), and demonstrate how light can enhance topological protection against lattice perturbations. Our work has potential practical implications for the ultrafast switching of materials properties, like optical band gaps or anomalous magnetoresistance. Moreover, we introduce Floquet time-dependent density functional theory (Floquet-TDDFT) as a general and robust first principles method for predictive Floquet engineering of topological states of matter.
\end{abstract}

\maketitle


\section{Introduction}

Creating and controlling phases of matter is a central goal of condensed matter physics. Recent advances in ultrafast spectroscopy\cite{orenstein_ultrafast_2012,zhang_dynamics_2014} open a route towards engineering new phases with pump laser fields acting on a solid to form emergent light-matter coupled states. As an example, circularly polarised light has been shown to open a band gap and produce Floquet-Bloch states
on the surface of an equilibrium topological insulator\cite{wang_observation_2013,mahmood_selective_2016}, and form a Floquet-Chern insulating state in graphene in the limit of continuous driving\cite{oka_photovoltaic_2009,kitagawa_transport_2011} and for finite pulse durations\cite{sentef_theory_2015}. 
The discovery of topological states in Dirac materials has triggered a lot of interest in particular in emergent Dirac\cite{novoselov_two-dimensional_2005}, Weyl\cite{nielsen_adler-bell-jackiw_1983}, and Majorana fermions\cite{kitaev_unpaired_2001,fu_superconducting_2008}. Topological states of matter are controlled by symmetries\cite{kitaev_periodic_2009,hasan_textitcolloquium_2010,qi_topological_2011}. Traditionally, the symmetries of materials can be influenced only to a certain extent and only on slow time scales via strain, doping, or static magnetic or electric fields. By contrast, Floquet engineering\cite{hanggi_driven_1998} allows to dynamically break symmetries and modify the topology of band structures\cite{lindner_floquet_2011} on ultrafast time scales.

Massless fermions in 3D Dirac and Weyl semimetals have recently attracted considerable interest. Examples include the Dirac semimetal materials Na$_3$Bi\cite{wang_dirac_2012,liu_discovery_2014} and Cd$_3$As$_2$\cite{wang_three-dimensional_2013,ali_crystal_2014,neupane_observation_2014,liu_stable_2014,liang_ultrahigh_2015,borisenko_experimental_2014,jeon_landau_2014}, and the Weyl semimetal states in transition-metal monophosphides \cite{weng_weyl_2015,huang_weyl_2015,liu_evolution_2016}, first discovered in TaAs\cite{xu_discovery_2015,lv_experimental_2015,lv_observation_2015,xu_discovery_2015-1,yang_weyl_2015,xu_observation_2016,xu_experimental_2015}. Besides the fundamental importance of Weyl semimetal materials as condensed matter realisations of elementary Weyl fermions, this interest is also due to the intrinsic stability of 3D Weyl points (WPs), which are chiral and host left- or right-handed Weyl fermions, 
giving rise to unusual material properties like negative magnetoresistance\cite{zyuzin_topological_2012,parameswaran_probing_2014,huang_observation_2015}, huge magnetoresistance\cite{shekhar_extremely_2015}, or the anomalous Hall effect\cite{yang_quantum_2011,burkov_weyl_2011,hosur_recent_2013,chan_when_2016}. WPs can alternatively be viewed as magnetic monopoles in momentum space with positive or negative chiral charges, and host non-zero Chern numbers for some closed momentum space surfaces
\cite{hasan_textitcolloquium_2010,qi_topological_2011}.

The topological protection of massless fermions in a Weyl semimetal against weak perturbations is controlled by the WP splitting in the Brillouin zone\cite{nielsen_absence_1981}, since the chiral WPs can only be destroyed by chirality mixing, which requires two opposite-chirality WPs to meet\cite{weng_weyl_2015}. The massless fermions in a Dirac semimetal, by contrast, require additional crystal symmetries to be stable, and are destroyed for instance by breaking rotational symmetry\cite{wang_dirac_2012}. Here, we propose a route towards ultrafast Floquet engineering of laser-induced topologically stable WPs starting from the 3D Dirac semimetal Na$_3$Bi by \textit{ab initio} electronic structure calculations using time-dependent density functional theory (TDDFT)\cite{runge_density-functional_1984,marques_fundamentals_2012}. We show that a Floquet-Weyl semimetal is dynamically created by breaking time-reversal symmetry. This symmetry breaking is achieved by applying circularly polarised classical laser fields with varying field strengths. Importantly, our strategy goes beyond the use of model tight-binding Hamiltonians and light coupling via Peierls substitution, since the TDDFT scheme automatically deals with the electronic properties and dynamical screening of the material and includes both Peierls phases for hopping terms and intra-atomic dipole (and higher multipole) transitions on equal footing\cite{bertsch_real-space_2000}.
In fact our theoretical framework shows effects that are not captured by simple four band models. For example, the splitting of degenerate Dirac bands into bands supporting a Floquet-Weyl point under a pump field does not occur symmetrically in all cases as would be predicted by models and under linearly polarised pumping the dynamical electron-electron interaction can induce a symmetry breaking field that destroys the Dirac point and opens a gap. 

\section{Theory}
To illustrate the basic idea behind the concept of  dynamically driven Floquet-Weyl semimetal we briefly discuss the minimal model in which Floquet-WPs arise from a 3D Dirac point and then move to a fully ab initio description of the real material Na$_3$Bi, where the dynamical electronic interactions of all valence electrons in the full Brillouin zone are taken into account. In a 3D Dirac semimetal, the Dirac point is a four-fold degenerate state at the Fermi level, and the low energy bands around this point obey a massless Dirac equation. Solutions to the massless Dirac equation are composed of two Weyl fermions with opposite chiralities, and the massless 4$\times$4 Dirac Hamiltonian can be written as a combination of uncoupled left- and right-handed 2$\times$2 Weyl Hamiltonians,
\begin{equation}\label{eq:Weyl}
    \hat{H}_{\rm Dirac}(\mathbf{k}) = \left(\begin{array}{cc}
        \hat{H}_{\rm Weyl}(\mathbf{k}) &  0  \\
         0 & \hat{H}_{\rm Weyl}(\mathbf{k})^* 
    \end{array}\right), \qquad H_{\rm Weyl}(\mathbf{k})= v_{\rm F} \mathbf{k}\cdot\mathbf{\sigma},
\end{equation}
where $v_{\rm F}$ is the Fermi velocity, $\mathbf{\sigma}=(\sigma_x,\sigma_y,\sigma_z)$ are Pauli pseudospin matrices, $\mathbf{k}$ measures momentum relative to the Dirac point, and $\hat{H}_{\rm Weyl}(\mathbf{k})$ is a right-handed Weyl Hamiltonian, while $\hat{H}_{\rm Weyl}(\mathbf{k})^*$ is its left-handed (time-reversed) partner, leading to eigenspinors with definite chirality. The 3D Dirac point thus consists of two degenerate WPs. In this case, the chiralities compensate each other because the WPs are at the same point in momentum space. Moreover, the WPs in a 3D Dirac semimetal are destroyed by any chirality-mixing perturbation that leads to a hybridisation of the subblocks in equation~(\ref{eq:Weyl}). By contrast, in a Weyl semimetal the left- and right-handed WPs are split in momentum space and are thus not susceptible to chirality-mixing perturbations. 

Starting from a 3D Dirac semimetal, see Fig.~1a, one can induce phase changes by lifting the fourfold degeneracy of the 3D Dirac point, either by introducing a mass term or by separating the degenerate WPs in momentum space. A mass term in the Dirac equation opens a gap and can stem from breaking rotational symmetry as induced by applying strain, leading to a topological insulator\cite{wang_dirac_2012}. The separation of the WPs into a Weyl semimetal state can be achieved by breaking time-reversal symmetry via an external driving, as shown in Fig.~1b. For the sake of the simplicity of the model but without loosing generality, we introduce the coupling to a time-dependent external gauge field by Peierls substitution in equation~(1), which amounts to $\mathbf{k}\rightarrow\mathbf{k}-\mathbf{A}(t)$, where $\mathbf{A}$ is the time-dependent vector potential of the applied circularly polarised light, e.g., $\mathbf{A}(t)=A_0(0,\cos(\Omega t),\sin(\Omega(t)))$ for polarisation in the $y$-$z$ plane. This treatment only takes into account the electric field of the laser pulse and neglects its magnetic component, which is negligible here but can also be straightforwardly included in the ab initio calculation.
The resulting time-dependent Hamiltonian $H(t) = H(\mathbf{k}-\mathbf{A}(t))$ describes the dynamics of the driven model system. From here on we suppress the momentum argument $\mathbf{k}$ for brevity.  

After transient effects have decayed, the system is in a stationary but non-equilibrium state of light-matter coupling that is periodic in time. Such a state can be analysed by Floquet theory, where the time dependence is described by mapping to a Hilbert space of time-independent multi-photon Hamiltonians, each projected onto a multiple of the photon frequency, effectively a Fourier-Bloch decomposition:
\begin{equation}\label{eq:Floquet_hamiltonian}
  \mathcal{H}^{mn}  = \frac{\Omega}{2 \pi}\int_T dt e^{i(m-n)\Omega t} H(t) + \delta_{mn}m\Omega \mathbf{1}
\end{equation}
where the integers $m$ and $n$ define the multi-photon Hilbert space (see Methods). This provides an interpretation of the time dependence in terms of time-independent multi-photon states with a well defined quasi-energy bandstructure, the Floquet bands. In the high-frequency limit one can decouple the zero-photon dressed states from the other states, amounting to a simple time-average, and add multi-photon states perturbatively\cite{oka_photovoltaic_2009,bukov_universal_2015}, 
\begin{equation}
  \hat{H}_{\rm eff} = \mathcal{H}^{00} +
  \frac{1}{\Omega}\left[\mathcal{H}^{0-1},\mathcal{H}^{01}\right] .
\end{equation}
where $\mathcal{H}^{00}$ is the zero order (cycle-averaged) Hamiltonian and $\mathcal{H}^{0\pm 1}$ are the single photon ``dressed'' Hamiltonians, c.f equation~(\ref{eq:Floquet_hamiltonian}). This Floquet downfolding has the advantage that one recovers the original Hilbert space of the electronic system and the resulting effective Hamiltonian $\hat{H}_{\rm eff}$ of this simple model retains an analytical form. Thus, for the the Dirac Hamiltonian, equation~(1), and $y$-$z$ circularly polarised light, the downfolded effective Floquet Hamiltonian reads 
\begin{equation}\label{eq:heff}
    \hat{H}_{\rm eff} = \left(\begin{array}{cc}
        \hat{H}_{\rm Weyl}(\mathbf{k}) + \frac{(v_{\rm F}A_0)^2}{\Omega}\sigma_x &  0  \\
         0 & \hat{H}_{\rm Weyl}(\mathbf{k})^* - \frac{(v_{\rm F}A_0)^2}{\Omega}\sigma_x 
    \end{array}\right) 
\end{equation}

The effective gauge field acts on the the $x$-component of the wave vector and results in a shift in the position of each of the originally degenerate WPs by $\pm (v_{\rm F}A_0)^2/\Omega$ along the $k_x$-direction in momentum space creating a Floquet-Weyl semimetal. This mechanism of light-induced effective gauge fields that alters the topology of the material is the central effect discussed in this paper. By solving the full ab initio mutli-photon Floquet-Hamiltonian without resorting to a high-frequency expansion or model parameters, we show that this effect occurs in real 3D Dirac semimetals and has the same dependence on amplitude and frequency but with a different prefactor. However, due to asymmetry of the bands in the actual material Na$_3$Bi it can also lead to a non-symmetric shift of the two WPs in energy away from the Fermi level, one upwards and the other one downwards, forming a topological Weyl metal. Thus by controlling amplitude, polarisation, and envelope of the driving field one can engineer different nontrivial Floquet topological phases. In particular, the same mechanism can be used to induce the Floquet-Weyl semimetal phase in systems that initially have a finite gap. Applying strain to a 3D Dirac semimetal opens a gap that can subsequently be closed by applying the driving field and with sufficient field strength the Weyl semimetal can be recovered, implying the possibility of Floquet-engineering a non-equilibrium metal-insulator transition.

For a realistic description of the 3D-Dirac semimetal Na$_3$Bi (see Fig.~1) and its insulating strained version under the influence of a periodic driving field, we use the first principles formalism of TDDFT\cite{marques_fundamentals_2012,runge_density-functional_1984,bertsch_real-space_2000,andrade_real-space_2015}. In this approach the electronic density is propagated in real time according to the time-dependent Kohn-Sham Hamiltonian, which accounts for electron-ion and electron-electron interactions through the time-dependent Hartree and the Kohn-Sham exchange and correlation potentials. The external gauge field is naturally included in the theory up to all orders as a time-dependent phase, as are the electronic multipoles to all orders. The time-dependent electronic structure from TDDFT is then further processed using the Floquet expansion(see Methods). This Floquet-TDDFT method represents a powerful and flexible tool to analyse and interpret time-dependent simulations, retaining both the rigorous \textit{ab initio} description of the electronic ground and time-dependent excited states while allowing for the readily accessible quasi-static picture of Floquet theory. The emergent Floquet bands can be viewed as snapshots of the electronic structure, time-averaged over the fast oscillation period of the external driving field, but time-resolved on the slower time scale of the pump pulse duration. This interpretation naturally reflects the measurement process in pump-probe photoemission spectroscopy\cite{freericks_theoretical_2009,sentef_theory_2015}. 

Both the carrier acceleration via Peierls substitution and the optical dipole transitions are automatically included in the TDDFT time propagation. 
For example, the observed asymmetric band splitting in the Weyl metal case would not follow from a simple tight-binding model but would require including dipole transitions that are a priori unknown for such a model and would therefore have to be retrofitted.
Moreover, by comparing the full TDDFT results including the exchange-correlation potential with a noninteracting time propagation, we find that the induced Hartree and exchange-correlation potentials do not play a crucial role for the effects described in this paper. Hence, the results are indeed generic. However, we stress that for different setups and effects beyond the ones discussed here, electron-electron interactions or electron-phonon interactions may indeed become important and will be accounted for by our theoretical approach. The Floquet-TDDFT framework introduced here will equally well apply to these situations and many more.

In the picture of photon-dressed states, Floquet sidebands are created as replicas of the original bands, spaced by the photon energy and periodically repeated in quasi-energy, in an enhanced Hilbert space including multiphoton processes. In practice, these replicas will be observed if the sidebands are occupied\cite{wang_observation_2013,sentef_theory_2015}, but they do not influence the physics discussed in this paper, in contrast to a different recent proposal for emergent Floquet-Weyl points at sideband crossings\cite{zou_floquet_2016}. Importantly, the main effect exploited in our work is due to off-resonant photon absorption, which is why the photon frequency is a freely tunable parameter that allows one to move the Floquet sidebands, for instance if sidebands from other bands further away from the Fermi level happened to interfere with the Floquet-Weyl points.

\begin{figure}
    \centering
     \resizebox{0.7\textwidth}{!}{\includegraphics{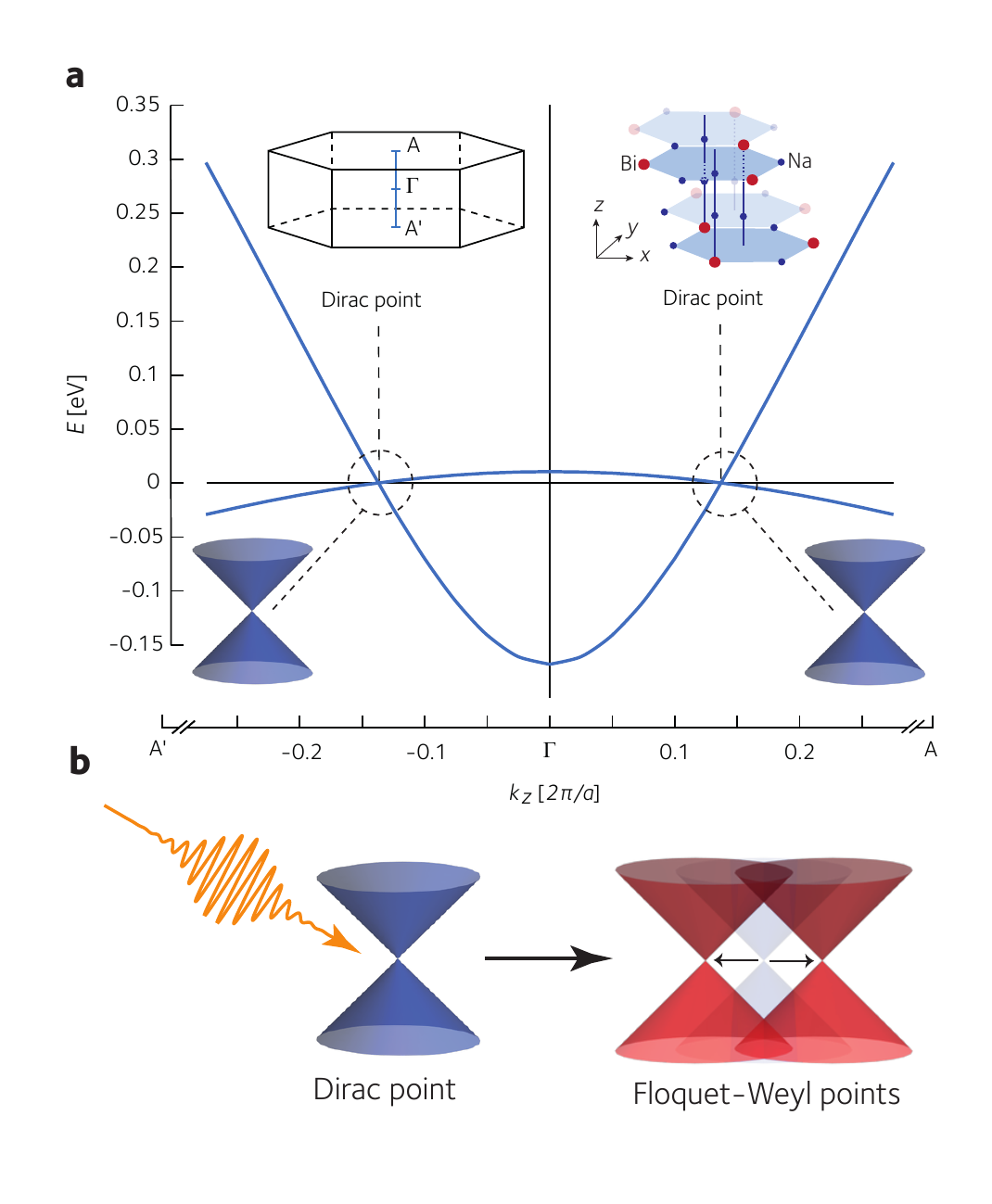}}
    \caption{{\bf Laser-induced Floquet-Weyl semimetal.} (a) With intact time-reversal symmetry, doubly degenerate 3D Dirac cones exists at two Dirac points in Na$_3$Bi. Shown are the Dirac bands along a section of the $A'-\Gamma-A$ path through the Brillouin zone, see left inset. The right inset shows the crystal structure of Na$_3$Bi. (b) Breaking time-reversal symmetry with circularly polarised light splits each Dirac cone into a pair of Floquet-Weyl cones.}
\end{figure}

\section{Floquet-Weyl points induced by circularly polarised light}

Applying time-reversal symmetry breaking fields to a 3D Dirac material lifts the degeneracy of the Dirac point into WPs leading to two distinct Floquet-Weyl cones emerging from each Dirac point in the Brillouin zone. In Fig.~2 we illustrate this effect using circularly polarised light with two different polarisation planes. While the $k_x$ and $k_y$ directions in the Brillouin zone of Na$_3$Bi are equivalent, and both showing symmetrical Dirac cones in the equilibrium phase, the dispersion of the Dirac cone along the $k_z$ direction is asymmetric\cite{wang_dirac_2012}, see Fig.~1a. This leads to qualitatively different behaviour of the driven system along the $k_z$ direction as opposed to the symmetric effect in the $x$-$y$ plane. Circularly polarised light in the $y$-$z$-plane splits the the Floquet-WPs along the $k_x$ direction forming a Floquet-Weyl semimetal, as shown in Fig.~2c. Conversely, with light polarised in the $x$-$y$ plane the Floquet-WPs split along the $k_z$ direction, cf.~Fig.~2f. Due to the asymmetry of the equilibrium Dirac cone along $k_z$ the Floquet-WPs are not only split horizontally but also shift vertically away from the Fermi energy. Crucially, one Floquet-WP moves to positive energies while the other one is shifted below the Fermi energy. The resulting complex intersection of the two Floquet-Weyl cones is shown in Fig.~2e. The shift in energy of the Floquet-WPs leads to the opening of two Fermi surfaces around these Floquet-WPs with opposing Chern numbers due to the opposite chiralities of the enclosed Floquet-WPs\cite{wang_dirac_2012}.

\begin{figure}
    \centering
     \resizebox{\textwidth}{!}{\includegraphics{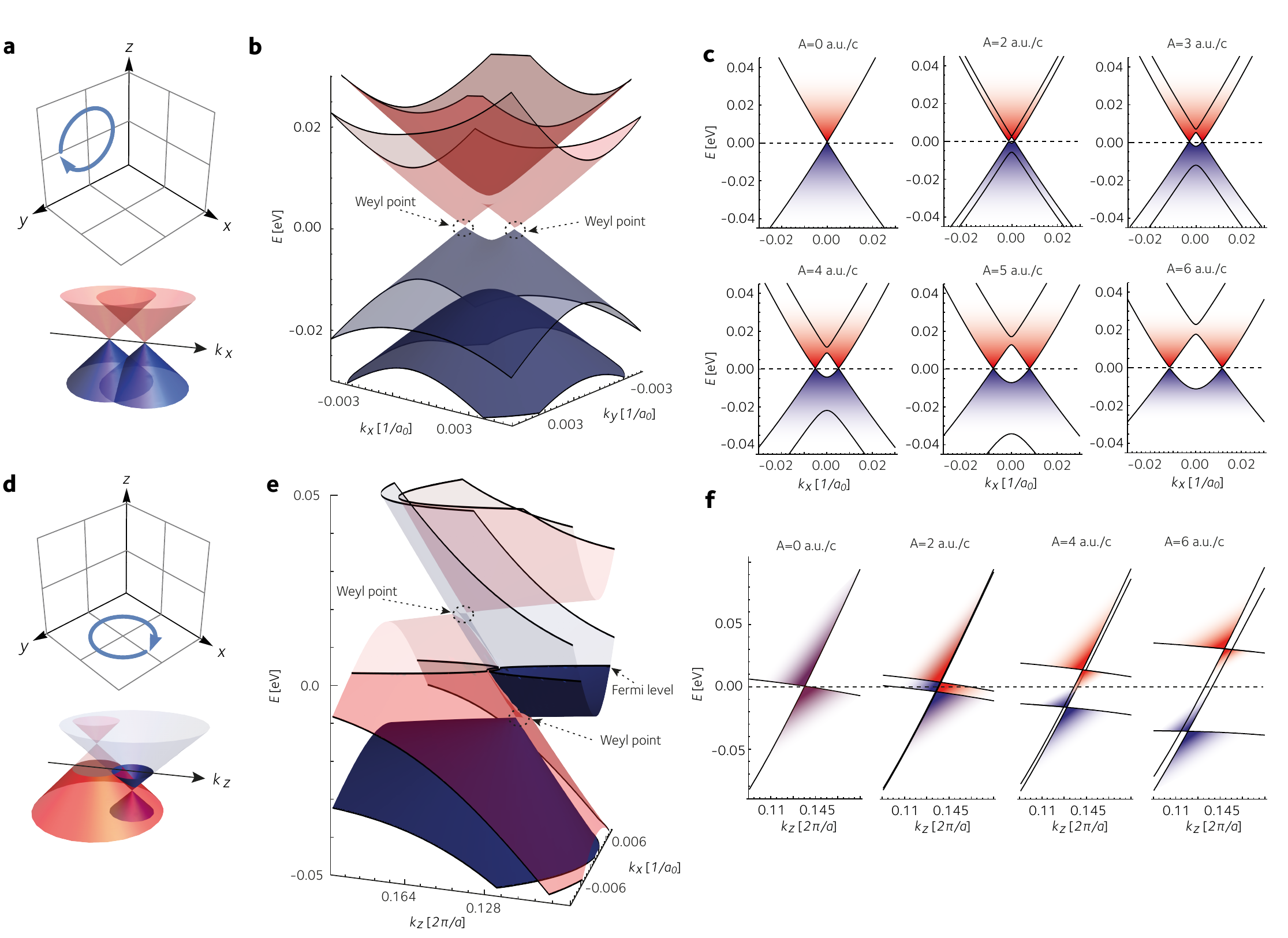}}
    \caption{{\bf Polarisation control of Floquet-Weyl point splitting.} (a) Circularly polarised light in the $y$-$z$-plane splits Floquet-WPs along the $k_x$ direction at the Fermi level, resulting in a Floquet-Weyl semimetal. (b) The Floquet-WPs form 3D cones in the Brillouin zone. An example in the $k_x$-$k_y$-plane is shown for amplitude $A=3$ a.u./$c$. (c) The splitting of the Floquet-WPs increases with the intensity (amplitude) of the driving field. Here we show the effect of different amplitudes $A$ of a field with photon energy $\hbar\Omega=1.5eV$ in momentum cuts around the Dirac point along $k_x$. The units of the amplitude of the vector potential can translated to electric field strength in units of V/\AA , see Methods. (d) With a similar driving field, but polarised in the $x$-$y$-plane,  the Floquet-WPs split along the $k_z$ direction, but in this case move away from the Fermi level, as shown in panel (f) for a series of amplitudes. (e) shows how this leads with an amplitude of $A=4$ a.u./$c$ to an opening of topological Fermi surfaces (rings in the $k_x$-$k_z$ plane) around the Floquet-WPs and a complex intersection of the Floquet-Weyl cones. In all panels, the Fermi level is at zero energy.}
\end{figure}

\section{Steering Floquet-Weyl points}
\begin{figure}
    \centering
     \resizebox{\textwidth}{!}{\includegraphics{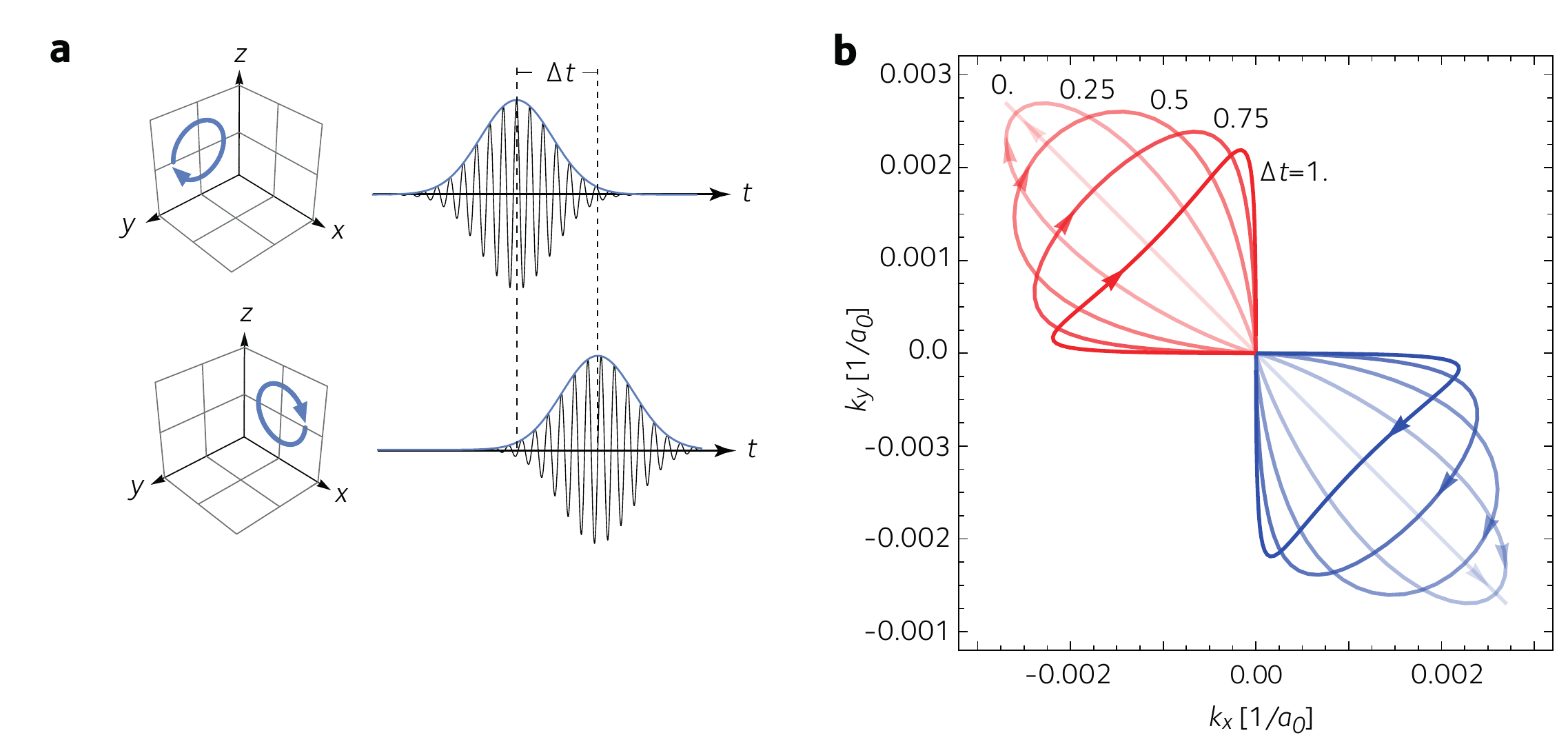}}
    \caption{{\bf Dynamical Floquet-Weyl points.} (a) Floquet-TDDFT predicts that the position of the Floquet-WPs can be controlled in real time on the time scale of pulse envelopes, yielding a time-dependent bandstructure that is accessible through time- and angle-resolved photoemission spectroscopy. Two time-delayed ($\Delta t$) pump laser pulses, circularly polarised in $y$-$z$ and $x$-$z$ directions, respectively, drive the Floquet-WP in the $k_x$-$k_y$ plane of the BZ. (b) The varying field strength during the laser pulses leads to ``dancing'' Floquet-WPs with trajectories controlled by $\Delta t$ given in units of the FWHM of pulse envelopes.}
\end{figure}

In a pump-probe experiment there are two different time-scales at play, the time of the period of oscillation and the time of the modulation of the amplitude due to the shape of the pump-pulse. The pump duration is typically orders of magnitude longer than the oscillation of the field and hence the pulse shape has no effect on the formation of Floquet-Weyl points at any given time during the pumping. In our calculations we observe that the Floquet limit is reached after two cycles of the driving field, and thus this assumption holds even for relatively short pulses of 100~fs. However, the changing of the envelope amplitude over time will change the position of the Weyl points according to Fig.~2. This means that on the time-scale of the pump-envelope there is a movement of the Weyl points through the Brillouin zone. Furthermore, from the previous discussion it has become clear that combining different pump polarisation offers the possibility to control the position of the Floquet-WPs within the Brillouin zone. Thus, by modulating the amplitudes of two pump fields, i.e. by controlling the delay between two pump pulses, one can time-dependently steer the Floquet-WPs. Fig.~3 shows how a simple time delay between two Gaussian pump pulses in the $x$-$z$ and $y$-$z$ planes controls complex trajectories of the two Floquet-WPs through the Brillouin zone. With more sophisticated combinations of different pump envelopes one can design complex dynamics of Floquet-WPs on the time scale of the pulse duration. Thus Floquet-TDDFT also provides a way of analysing time-dependent band structures, that are different from the instantaneous eigenvalues of the TDDFT Hamiltonian, and that can be measured using time-resolved photoemission spectroscopy, which has an intrinsic time-averaging procedure given by the probe pulse shape\cite{sentef_theory_2015}. We also remark here that in principle further replicas of the dancing Floquet-WPs may be observed as an effect of Floquet sideband formation.

\section{Strain}

\begin{figure}
    \centering
    \resizebox{\textwidth}{!}{\includegraphics{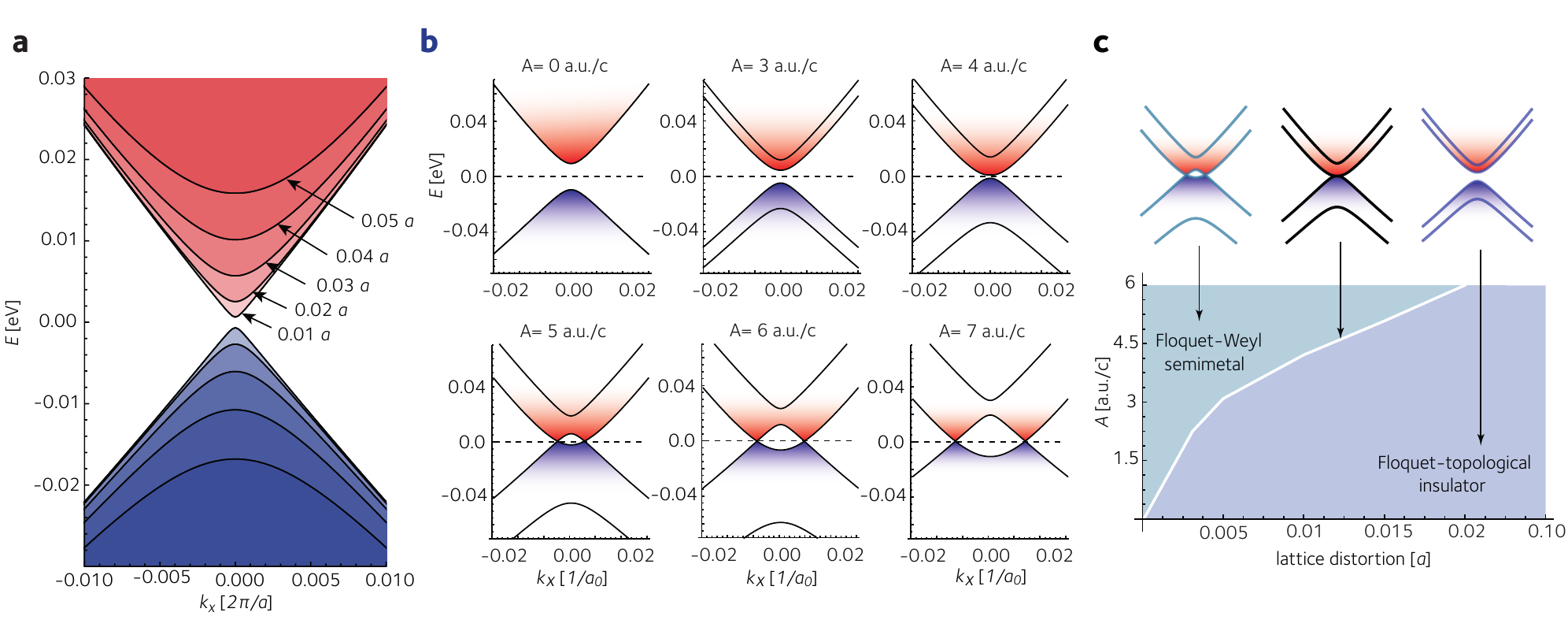}}
    \caption{{\bf Floquet topological phase transition and enhanced topological protection.} (a) In strained samples, the breaking of crystal symmetries opens a gap at the 3D Dirac point, leading to a topological insulator phase. In Na$_3$Bi, strain along $x$ breaks the $C_3$ symmetry. Shown is a series of bandstructures for different lattice distortions measured as a fraction of the lattice parameter. (b) The strain-induced band gaps can be closed by the application of circularly polarised light. For 1\% lattice distortion, increasing the amplitude of a $y$-$z$ circularly polarised driving field with $\hbar\Omega=1.5$eV first reduces the gap, then closes it at a critical amplitude $A_c$, and restores the Floquet-Weyl semimetal observed in the unstrained case (see Fig.~2c). The units of the amplitude of the vector potential can translated to electric field strength in units of V/\AA. (c) This light-induced Floquet topological phase transition from topological insulator to Weyl semimetal defines a non-equilibrium phase diagram as a function of amplitude and lattice distortion for a given driving field frequency.}
\end{figure}

In all Dirac metals the doubly degenerate 3D Dirac point of the equilibrium phase of the material leaves the topology unstable under small lattice deformations. In Na$_3$Bi, already a small strain breaks the rotational $\mathcal{C}_3$ symmetry and leads to an opening of a gap due to chirality mixing and the formation of a topological insulator phase\cite{wang_dirac_2012}, Fig.~4a. The Floquet-WPs, however, are protected by conservation of chirality and afford the system topological protection. In fact, for an initially gapped system, the application of the driving field can lead to a restoration of the topology and the Floquet-Weyl semimetal phase. In Fig.~4b the progression from the strained gapped topological insulator phase to the Floquet-Weyl semimetal phase is shown as the amplitude of the driving field increases. The transition from one phase to the other occurs at a critical amplitude that characterises a phase boundary in a Floquet phase diagram, see Fig.~4c. We note that  a related idea was used to drive a topological insulator into a Weyl semimetal phase in Ref.~\onlinecite{wang_floquet_2014}.

Besides being a striking illustration of the power of topological protection this light-induced phase transition might also be of technological relevance, because it facilitates light-controlled, ultrafast switching between an insulating and metallic phase. In particular when the driving field is circularly polarised in the $x$-$y$ plane the topological insulator phase can be switched to a phase with non-trivial topological Fermi surfaces, possibly allowing for a technological exploitation of the exotic transport phenomena encountered in this phase\cite{huang_observation_2015,shekhar_extremely_2015}. 

\section{Discussion}
Our work demonstrates how topological properties in solids can be Floquet-engineered. Floquet-TDDFT predicts distinct Weyl semimetal, Weyl metal, and topological insulator Floquet band structures that can be measured with time-resolved photoemission spectroscopy. The splitting of Floquet-Weyl points controlled by laser driving illustrates how the concept of separation of chiral particles in momentum space inspires the Floquet-engineering of topologically stable states. Regarding anomalous transport properties, we envision that THz magnetotransport might be able to detect effects of Floquet-Weyl fermions near the Fermi level, for instance the anomalous negative magnetoresistance corresponding to the chiral Adler-Bell-Jackiw anomaly\cite{zyuzin_topological_2012,parameswaran_probing_2014,huang_observation_2015}, the recently demonstrated chiral magnetic effect\cite{li_chiral_2016}, a proposed chiral pumping effect with an axial current\cite{ebihara_chiral_2015}, or huge magnetoresistance for the Floquet-Weyl metal with hole and electron Fermi surfaces\cite{shekhar_extremely_2015} arising for $x$-$y$ polarisation. Our predictions can be directly checked in pump-probe angle-resolved photoemission experiments, which can measure the transient Floquet band structures\cite{wang_observation_2013,mahmood_selective_2016}. These experiments could additionally even provide evidence about topological features implied by topological insulator, Dirac semimetal and Weyl semimetal materials, e.g. Fermi arc surface states of Weyl semimetal or surface states in topological insulators. Furthermore, a recent work\cite{kolodrubetz_non-adiabatic_2016} proposes measurement of the Wigner distribution as a means to identify non-trivial Floquet phases.

In terms of method development, we have introduced the framework of Floquet-TDDFT as a tool for the first principles prediction of Floquet-engineered bands and topologies. Furthermore it provides a setting to discuss time-dependent band structures on time scales accessible in pump-probe experiments. Finally, electron correlation effects beyond independent particles in driven systems are thus within reach, paving the way for the study of nonlinear laser field and collective excitation effects, such as the coupling to phonons, or exciton and plasmon formation in realistic Floquet-driven complex materials.

\section{Acknowledgemnts}
We are grateful to Peizhe Tang for stimulating discussions and a critical reading of our manuscript, and to Ashvin Vishwanath for helpful discussions. We acknowledge financial support from the European Research Council (ERC-2015-AdG-694097), Grupos Consolidados (IT578-13), Spanish grant (FIS2013-46159-C3-1-P),
AFOSR Grant No. FA2386-15-1-0006 AOARD 144088, COST Action MP1306 (EUSpec). H.H. acknowledges support from the People Programme (Marie Curie Actions) of the European Union's Seventh Framework Programme FP7-PEOPLE-2013-IEF project No. 622934. A.F.K. would like to thank the Aspen Center for Physics (supported by National Science Foundation grant PHY-1066293) for their hospitality during part of this work.

\section{Methods}

\subsection{Units of field amplitudes}
The amplitudes of the vector potentials discussed in the text can be converted to electric field strength as $E=-\partial A/\partial t\sim -A\Omega$ assuming that the envelope is constant. This leads to a conversion of 1 [a.u./c] = $0.013789$ V/\AA $\Omega$ eV. For the frequency of $\hbar \Omega=1.5$ eV used here this corresponds to a conversion of all amplitudes $A$ given in the text as 1 [a.u./c] $=$ $0.0206849$ V/\AA $=$ $2.068498$ MV/cm.

\subsection{Floquet-TDDFT}
According to the Floquet theorem, the solutions of the time-dependent Schr\"odinger equation $i\partial_t |\psi(t)\rangle=\hat{H}(t)|\psi(t)\rangle$ with a Hamiltonian that is periodic in time, i.e. $\hat{H}(t+T) =\hat{H}(t)$, can be obtained by computing the eigenstates of a static Hamiltonian in the Hilbert space of multi-photon dressed states. This static Hamiltonian follows from performing a Floquet-Bloch expansion of the time-dependent solutions of the Schr\"odinger equation $|\psi(t)\rangle=\sum_{m}\exp(-i(\epsilon+m\Omega)t)|u_m\rangle$, where $\Omega=2\pi/T$, $\epsilon$ is the Floquet quasi-energy, and $|u_m\rangle$ is the corresponding $m$-th Floquet eigenfunction, which does not depend on time. The time-dependent Schr\"odinger equation then reduces to the static equation
\begin{equation}
    \sum_{n}\mathcal{H}^{mn} |u_n\rangle = \epsilon |u_m\rangle    
\end{equation}
for each $\epsilon$, where the $\mathcal{H}^{mn}$ is a static Hamiltonian
\begin{equation}
  \mathcal{H}^{mn}  = \frac{1}{T}\int_T dt e^{i(m-n)\Omega t} H(t) +
  \delta_{mn}m\Omega \mathbf{1}
\end{equation}
defined in the infinite Hilbert space of multi-photon (i.e. multiples of $\Omega$) components. Thus Floquet theory offers a way of analyzing periodically driven systems. In principle, the full description of the system requires the diagonalisation of the full Floquet-Hamiltonian while in practice one truncates the photon number depending on the problem at hand. In this work we found that the contributions of two-photon terms and beyond had a negligible effect on the bands considered here.

The motivation to perform Floquet analysis is to project the time dependence of the driven system to a static picture. While this is usually used to obtain an analytical expression that exposes the physical mechanism, we use it here literally as a tool to analyze time-dependent data. TDDFT gives the real-time evolution of the electronic density by propagating the Kohn-Sham states under any kind of static or time-dependent perturbation. The time evolution operator is built from the Kohn-Sham Hamiltonian $H_{\rm KS}$ with explicitly time-dependent external fields $V_{\rm ext}$:
\begin{equation}
  H_{\rm KS}(t) = T+V_0+V_{\rm H}[n(t)]+V_{xc}[n(t)]+V_{\rm ext}(t)
\end{equation} 
where $T$ and $V_0$ are the kinetic energy and static potential, while the Hartree potential $V_H$ and the exchange and correlation potential $V_{xc}$ dynamically depend on the density $n(t)$ during the time evolution. For extended systems we use the velocity gauge concept introduced in Ref.~\onlinecite{bertsch_real-space_2000} to treat, within the dipole approximation, the response of extended periodic systems to an arbitrary time-dependent perturbation. Within the velocity gauge the external potential arises in the Hamiltonian from the substitution $\mathbf{p}\rightarrow \mathbf{p}-\mathbf{A}(t)/c$ leading to the terms $1/2(\mathbf{p}-\mathbf{A}(t)/c)^2=T+1/2(A(t)/c)^2 -\mathbf{p}\cdot\mathbf{A}(t)/c = T+V_{\rm ext}(t)$ (in atomic units).

Besides the fact that the physical properties of the system over time can then be derived from the time-dependent density, it is obvious that this approach also generates a Hamiltonian at each time step. If the system is driven by a periodic external field or finds itself in an otherwise periodically oscillating state, such as phonon modes, this Hamiltonian fulfills the Floquet condition of periodicity and can be used directly in Eqs.~(\ref{eq:Floquet_hamiltonian}). Hence, Floquet analysis provides an approach of processing real-time propagation results from TDDFT to obtain spectral information that is richer than the instantaneous Kohn-Sham eigenvalues. 

In practice, the Floquet-TDDFT requires storing the time-dependent Hamiltonian, not just its instantaneous eigenvalues, during one cycle of periodicity so it can be used in the time integral of equation~(\ref{eq:Floquet_hamiltonian}). This would require an unfeasible amount of storage for most systems in a real-space or wavevector basis. Instead, it is sufficient to save either the time-dependent density or the implicitly time-dependent potentials $V_{\rm xc}$ and $V_{\rm H}$. It is also worth noting, that since the evolution of the system is smooth, the time integrals in equation~(\ref{eq:Floquet_hamiltonian}) can be evaluated with a coarse time sampling of the Hamiltonian. Furthermore, the Floquet condition of periodicity does not have to be fulfilled \textit{a priori} but one can perform the Floquet analysis over any time interval corresponding to the cycle of periodicity and determine the Floquet limit as a convergence of the Floquet bands with propagation time.

We note that in the Floquet-TDDFT results, we always observe a small shift of the energies of maximum $20$meV for large amplitudes, that we assign to numerical inaccuracies. This shift does not change the character of the Weyl semimetal phase, because it just corresponds to the case of a doped Weyl semimetal, and therefore we have compensated it in the all the figures.

The fully ab initio nature of Floquet-TDDFT means that it requires no fitting reference. Instead, it might indeed be used as a reference itself for simplified qualitative tight-binding models that reduce the observed effect to its essential ingredients. For example one might think of using such an approach to refine the illustrative model we use in Eq.~(\ref{eq:Weyl}). While this can be useful in some case one, however looses the quantitative accuracy that TDDFT provides and in this case leads to quantitatively wrong dynamics of the Weyl points.

\subsection{Computational details}
We carry out DFT and TDDFT calculations as implemented in the OCTOPUS code\cite{andrade_real-space_2015} using the (adiabatic) local density approximation LDA and HGH pseudopotenitals\cite{hartwigsen_relativistic_1998} including spin-orbit coupling. The structure is taken from Ref.~\onlinecite{wang_dirac_2012} where the lattice parameter is given as $a=10.2952271$ Bohr. The groundstates are computed with a $10\times 5\times 5$ Monkhorst-Pack Brillouin zone sampling and a real-space mesh of $0.3$ Bohr spacing. The TDDFT calculations are performed with a time step of $0.05$ $\hbar$/Ha and with a uniform external vector potential oscillating with a frequency of $\hbar\Omega=1.5$eV and with different amplitudes specified in the text. We find that for our calculations of Na$_3$Bi the Floquet bandstructures do not depend on the LDA functional and also the effect of the induced Hartree potential is negligible for circular polarised driving fields. For the Floquet analysis, the Hamiltonian is saved at 70 equally spaced points in time during one cycle and then processed as described in the Floquet-TDDFT section.

\bibliography{NaBi_topological}


\end{document}